\documentclass[aps]{revtex4}

\usepackage{amssymb,amsmath,graphicx}
\begin{document}
\title{Probabilistic whereabouts of the "quantum potential"}
\author{Piotr Garbaczewski}
\address{Institute of Physics, University of Opole, 45-052 Opole, Poland}
\begin{abstract}
We review major appearances of the functional expression $\pm \Delta \rho ^{1/2}/ \rho ^{1/2}$
 in the theory of diffusion-type processes  and in  quantum mechanically supported dynamical scenarios.
Attention is paid to various manifestations of "pressure" terms  and their meaning(s)   in-there.
  \end{abstract}
\maketitle

\section{From  $\rho $  to  $\pm {\frac{ \Delta {\rho }^{1/2}}{\rho ^{1/2}}}$.}

For clarity of presentation, we restrict considerations to  probability density functions (pdfs) in one spatial dimension, with the
 time label suppressed for a while (albeit we have in mind not only random variables, but  stochastic processes as well).
  Given  a continuous, at least twice differentiable  pdf  $\rho (x)$,     $\int_R \rho (x) \, dx =1$  we  infer
 a hierarchy of   functions   $ - \ln \rho (x)$, $-\nabla \ln \rho $ and  $-\Delta \ln \rho $ whose meaning will soon become more  transparent.

An information theory notion  of the  Shannon  entropy of  $\rho $  (with varied interpretations, among which the notions  of disorder and/or uncertainty
seem to prevail)  reads
\begin{equation}
{\cal{S}}(\rho )=  - \langle \ln \rho \rangle = - \int \rho (x) \ln
\rho (x)  dx
\end{equation}
thus giving $\ln \rho $  the meaning of (Shannon) entropy density.
An affiliated measure of disorder/uncertainty  named the  Fisher information   of $\rho $  has the form
\begin{equation}
{\cal{F}} (\rho ) \doteq  \langle ({\nabla \ln \rho })^2\rangle =
\int {\frac{(\nabla \rho )^2}{\rho }}  dx
\end{equation}
and in view of  $\langle \nabla \ln \rho \rangle =0$,  stands for a mean square deviation  of  a function   $\nabla \ln \rho (x) $
of the random variable $X$, with values $x\in R$.

These  two information theory measures are interrelated, as for example can be  seen through so-called   isoperimetric
 inequality: ${\cal{F}} \geq (2\pi e) \exp(2{\cal{S}})$. More than that, it is the Fisher information which is directly involved in the primordial
 form of the indeterminacy (uncertainty) principle.  Namely,  $\langle \nabla \ln x  \rangle =0$ and
 $Var(x)=\sigma ^2 = \langle (x- \langle x\rangle )^2\rangle $  imply   an  indeterminacy relationship (no quantum context as yet):
\begin{equation}
Var(\nabla \ln x) = {\cal{F}} (\rho ) \geq 1/\sigma ^2>0 \, .
\end{equation}
We can get  a  deeper insight  into the indeterminacy rule, by  noting that  actually  $\rho ^{1/2}$ is a square integrable function. Therefore,
 standard $L^2(R)$ Fourier transform techniques $\psi \rightarrow \tilde{\psi }$   can be here  adopted.  In a slightly  more general  notation,
  which encompasses the previous case,
 $\rho = |\psi |^2,  \psi \in L^2$ $ \Longrightarrow$ $\tilde{\psi }$, $\tilde{\rho } = |\tilde{\psi }|^2$,   one  infers (in self-explanatory notation)
\begin{equation}
(1/\sigma ^2) \leq {\cal{F}} \leq 16 \pi ^2 \tilde{\sigma}^2
\end{equation}
and
\begin{equation}
(4\pi /\tilde{\sigma}) \leq (1/ \sqrt{2\pi e)}  \exp [{\cal {S}}] \leq \sigma .
\end{equation}

We can continue our discussion of $ - \ln \rho (x)$, $-\nabla \ln \rho $ and  $-\Delta \ln \rho $     by  simply playing  with   them, to
reveal  a number of  emergent quantities, like  e.g.
\begin{equation}
- \Delta  \ln \rho = -
{\frac{\Delta \rho }\rho } + {\frac{(\nabla \rho )^2}{\rho ^2}}  \Longrightarrow
 -\langle \Delta \ln \rho \rangle = \langle
{\frac{(\nabla \rho )^2}{\rho ^2}}\rangle = {\cal{F}} (\rho )   \, .   \label{por}
\end{equation}

At this point we   introduce a potential  $\frac{\Delta \rho ^{1/2}}{\rho ^{1/2}}$    of  a    Newton-type  "force field" with
 vanishing mean  value   $\langle \nabla ({\frac{\Delta \rho ^{1/2}}{\rho ^{1/2}}}) \rangle = 0 $
 \begin{equation}
- {\frac{\Delta \rho ^{1/2}}{\rho ^{1/2}}}  = {\frac{1}2}
[- {\frac{\Delta \rho }\rho }  + {\frac{1}2} {\frac{(\nabla \rho )^2}{\rho ^2}}] \, \, \, \Longrightarrow  \, \, \,
+  \nabla ({\frac{\Delta \rho ^{1/2}}{\rho ^{1/2}}}) =   {\frac{1}{2\rho }}   \nabla (\rho \Delta \ln \rho )  \label{fisher0}
\end{equation}
and make explicit  its links with the Fisher information  of the pdf   $\rho $, through
\begin{equation}
-  \langle  {\frac{\Delta \rho ^{1/2}}{\rho ^{1/2}}} \rangle =
   - {\frac{1}4}   \langle \Delta \ln \rho
\rangle = {\frac{1}4}{\cal{F}}(\rho ) \geq {\frac{1}{4 Var(x)}}>0  \, ,
\end{equation}
We emphasize a conspicuous absence  of any   specific  physical context. Nonetheless,  while accounting for a temporal evolution
of $\rho = \rho (x,t)$,  a number of  physically interesting quantities can be easily  identified. They are
   omnipresent in local conservation  laws for diffusion-type stochastic  processes, as well as  in  the
  hydrodynamical  formulation of the  Schr\"{o}dinger picture  quantum dynamics.

\section{Emergence of $\pm {\frac{ \Delta {\rho }^{1/2}}{\rho ^{1/2}}}$ in hydrodynamical (local) conservation laws}

\subsection{Quantum hydrodynamics}
Taking as obvious the standard wisdom about a hydrodynamical representation of the Schr\"{o}dinger picture quantum dynamics, we merely recall that the Schr\"{o}dinger equation
\begin{equation}
  i\hbar \partial _t \psi  = \left[ - {\frac{\hbar ^2}{2m}} \Delta +
V \right] \psi
\end{equation}
involves a Hamiltonian
$\hat{H}$  that is  self-adjoint  operator in a suitable Hilbert space domain.   Since we shall be dealing with
bounded from below operators,  for later convenience we  impose an additive
renormalization of the Hamiltonian so that
 $\hat{H} \geq 0$.   Further notations  are reproduced  for the record:  the pdf is  $\rho (x,t)= |\psi |^2(x,t)$,
$v= (\hbar /2mi)[ (\nabla \psi /\psi ) -  (\nabla \psi ^*/\psi ^*)]$ stands for the current velocity field. With the
 polar (Madelung) decomposition of $\psi $ being implicit, we get:
\begin{equation}
\partial _t\rho = - \nabla (\rho \,   v); \, \,   \, \,  \, \, \, \, \partial _ts  +   {\frac{1}{2m}} (\nabla s)^2 + (V +  Q)
  = 0  \Longrightarrow
 \end{equation}
  $$  \partial _t v + (v \nabla v) = - {\frac{1}m} \nabla (V+Q)$$
where $v= {\frac{1}m} \nabla s$ \,  and   $Q  = Q[\rho ]=
  - {\frac{\hbar ^2}{2m}}\, {\frac{\Delta \rho ^{1/2}}{\rho ^{1/2}}} $  has a folk  name of  a "quantum potential".  \\
Set $|\psi |= \rho_*^{1/2}$.   The ground state condition  for $\hat{H}$, with bottom eigenvalue $0$,
 directly involves the "quantum potential":
  \begin{equation}
   V=+{\frac{\hbar ^2}{2m}}\, {\frac{\Delta \rho _*^{1/2}}{\rho _*^{1/2}}}  =-Q[ \rho _*]. \label{comp}
   \end{equation}
Denote $u(x,t) \doteq (\hbar / 2m)\, \nabla \ln \rho   $. It is well known that the
dynamics  arises via the   $\{\rho, s\}$   extremum principle for
 \begin{equation}
 I(\rho, s)= \int_{t_1}^{t_2}
 \langle \left[ \, \partial _ts      + {\frac{m}{2}} (u^2 + v^2)\,    +  V \right] \rangle (t) dt .
 \end{equation}
 In terms of  valid  solutions $\rho (x,t)$, $s(x,t)$, we arrive at a strictly positive   constant of motion:
        $-\langle \partial _t s \rangle = H= \langle \left[ {\frac{m}{2}} (u^2 + v^2)\,    +  V \right] \rangle >0$  (a finite energy condition).

 \subsection{Brownian hydrodynamics}

 The semigroup dynamics and the emergent generalized diffusion equation (note  that by setting  $V=0$ we pass to the standard heat equation)
 \begin{equation}
\exp(-t\hat{H}/2mD)\Psi _0 = \Psi _t \, \,  \Longrightarrow \, \,   \partial _t \Psi = \left[ D \Delta  -  {\frac{V}{2mD}}\right] \Psi
\end{equation}
is a  self-adjoint  relative of the  more familiar  Fokker-Planck equation $\partial _t\rho = D\Delta \rho - \nabla (b\rho )$ and likewise,
although   indirectly, determines the evolution of $\rho (x,t)$.
 Here $\hat{H}$  is  self-adjoint, $\hat{H} \geq 0$, $t\geq 0$. (We keep in mind  a re-definition  $\hbar \equiv 2mD$.)

 Let $\Psi (x,t) \rightarrow \rho _*^{1/2}$ as $t\rightarrow \infty $. Define $\rho (x,t)=\Psi (x,t) \rho _*^{1/2}(x)$ with
 $ b= D\nabla \ln \rho _* $, $ u= D\nabla \ln \rho $  and   $ v= b- u = (1/m)\nabla s $. The connection between the Fokker-Planck and
 semigroup dynamics is  being  established, provided a  compatibility condition
 \begin{equation}
 {\frac{V(x)}{2mD}} = + D {\frac{\Delta \rho _*^{1/2}}{\rho _*^{1/2}}} \doteq mD\left[ {\frac{b^2}{2D}} + \nabla b\right]
 \end{equation}
 holds true.
The  (rescaled)  "quantum potential"  appears  in the above (c.f. also \eqref{comp}), as well as  in Hamilton-Jacobi type equations  of motion:
\begin{equation}
\partial _t\rho = D\Delta \rho - \nabla (b\rho )  \Longleftrightarrow \partial _t\rho =
- \nabla (v \rho )
\end{equation}
$$\partial _t s + (1/2m) (\nabla s)^2  - (V + Q)=0 \, \,  \Longrightarrow  \, \,   \partial _t v + (v \nabla v) = + {\frac{1}m} \nabla (V+Q)  $$

 The $\{\rho, s\}$ extremum  principle  for
 \begin{equation}
I(\rho, s)= \int_{t_1}^{t_2}  \langle \left[ \, \partial _ts
      + (m/2)(v^2  -  u^2) - V\right]\rangle
\end{equation}
       yields the previous Hamilton-Jacobi type  dynamics.
       In terms of  dynamically admitted fields  $\rho (x,t)$ and $s(x,t)$, we   have
$-\langle \partial _t s\rangle  = H = \langle \left[{\frac{m}2}( v^2 - u^2)    -
 V  \right]\rangle  \equiv 0 $.

\subsection{Time dependence    of Shannon and Fisher functionals}

The dynamics of $\rho (x,t)$ is dictated by the continuity equation and this equation alone
sets the  evolution rule for the Shannon entropy  ${\cal{S}}[\rho ](t)$. Indeed, there holds (provided the fall-off of $\rho $ at spatial
infinities  ensures the  vanishing of $v\rho$):
\begin{equation}
 {\frac{d\cal{S}}{dt}}  = \langle \nabla v\rangle =  -  {\frac{1}D}\, \langle v\, u \rangle
\end{equation}
where  $D\equiv \hbar/2m$ corresponds to the quantum case. Obviously, there is an evolution of the velocity field $v(x,t)$ to be accounted for,
c.f. the corresponding Hamilont-Jacobi type equations and their gradient versions.

By exploiting the Hamilton-Jacobi type equations,  in case of time independent external potentials, we easily demonstrate that
\begin{equation}
 {\frac{d\cal{F}}{dt}} =  \mp 2 \langle v \nabla P \rangle
\end{equation}
where $P= \rho \Delta \ln \rho $. The minus sign  in the above  refers to diffusion processes, while the plus sign  to  the quantum motion.

A more detailed discussion of the pressure-type  term $P$ and its  functioning  will be given later.  We note a clear parallel with  a  classical power
release expression $dE/dt=v\, F$,   where $F=-\nabla V$ is a standard Newtonian force.

\section{\textbf{Dynamical duality - illusion of "Euclidean time"}}
In the light of our  previous discussion  there appears   quite  persuasive   to    execute (in the least formally) the Wick rotation
in the complex time plane $it \rightarrow t\geq 0;   \, \, \, \,  \hbar \rightarrow 2mD$
\begin{equation}
  \exp(-i\hat{H}t/\hbar) \psi _0 = \psi _t  \Longrightarrow
\exp(-t\hat{H}/2mD)\Psi _0 = \Psi _t
\end{equation}
that maps between diffusion-type and quantum mechanical patterns of dynamical behavior.
Given the spectral solution for $\hat{H}= - \Delta + V$,  the  integral kernel of   $\exp(-t\hat{H})$
reads \begin{equation}
k(y,x,t)= \sum_j \exp(- \epsilon _j t) \, \Phi _j(y) \Phi ^*_j(x).
\end{equation}
 Remember that $\epsilon _0=0$  and
the sum may be replaced by an integral in case of a continuous spectrum, (with complex-valued  generalized   eigenfunctions).
Set $V(x)=0$ identically. Then we end up  with a familiar heat kernel:
\begin{equation}
k(y,x,t)= [\exp(t\Delta )](y,x) = (2\pi )^{-1/2} \int \exp(-p^2t)\, \exp(ip(y-x)\, dp=
\end{equation}
$$(4\pi t)^{-1/2}\, \exp[-(y-x)^2/4t]$$

Consider $\hat{H}= (1/2)(-\Delta + x^2  - 1)$ (e.g. the  rescaled harmonic oscillator Hamiltonian). The integral kernel of $\exp(-t\hat{H})$
  is given by the classic   Mehler formula:
  \begin{equation}
 k(y,x,t) = k(x,y,t)= [\exp(-t\hat{H})(y,x)=
\end{equation}
  $$[\pi (1-\exp(-2t))^{-1/2} \exp[-(1/2)(x^2-y^2) - (1-\exp(-2t))^{-1}\, (x\exp(-t) - y)^2]$$
 The normalization condition  $\int k(y,x,t) \exp[(y^2-x^2)/2]\, dy =1$   actually  defines the transition probability density
 of the Ornstein-Uhlenbeck process
 \begin{equation}
p(y,x,t) = k(y,x,t)\, \rho _*^{1/2}(x)/ \rho _*^{1/2}(y)
\end{equation}
   with $\rho _*(x)=\pi ^{-1/2} \exp(-x^2)$.
 A more familiar form of the kernel reads (note the presence of $\exp(t/2)$ factor)
 \begin{equation}
 k(y,x,t) = {\frac{\exp(t/2)}{(2\pi  \sinh t)^{1/2}}}  \exp  \left[ - {\frac{(x^2+y^2)\cosh t  - 2xy}{2\sinh t }} \right]
\end{equation}
To conform with the statistical physics lore of the $50$-ies and $60$-ties, we   can easily pass  to an integral kernel of the density
operator, labeled by equilibrium values of the temperature. To this end one should  set e.g.  $t\equiv \hbar\omega /k_BT$ for a
harmonic  oscillator with a proper   frequency $\omega $ and
 remember about evaluating the normalization factor $1/Z_T$ where $Z_T$ stands for a partition function of the system.

Concerning the "Euclidean issue", we  note that by
formally executing  $t\rightarrow it$  one  arrives at the   free Schr\"{o}dinger   propagator
\begin{equation}
K(y,x,t)= [\exp(it\Delta )](y,x) = (2\pi )^{-1/2} \int \exp(-ip^2t)\, \exp(ip(y-x)\, dp=
\end{equation}
$$(4\pi it)^{-1/2}\, \exp[+i(y-x)^2/4t]$$
and likewise, at that of  (here $-1$  renormalized)  harmonic oscillator  propagator
\begin{equation}
 K(y,x,t) = {\frac{\exp(it/2)}{(2\pi i \sin t)^{1/2}}}  \exp  \left[ +i {\frac{(x^2+y^2)\cos t  - 2xy}{2\sin t }} \right]
 \end{equation}
 Learn  a standard Euclidean   (field) theory lesson concerning multi-time correlation functions. In the
  exemplary harmonic oscillator case, $t>t'>0$;  $t\rightarrow it$ results in:
\begin{equation}
E[X(t')X(t)] = \int  \rho _*(x')\,   x'\,  p(x',t',x,t)\,  x\,  dz dx'  = (1/2)\, \exp [-(t-t')]   \Longrightarrow
\end{equation}
 $$W(t',t) = \langle \psi _0, \hat{q}_H(t) \hat{q}_H(t')\psi _0\rangle =  (1/2) \exp [-i(t-t')]$$
where $\hat{q}_H(t)$ stands for the postion operator in the Heisenberg picture.
In passing, we note that this appealing correspondence breaks down in $R^n$, $n>1$,   in the presence of electromagnetic fields.

The major message of our discussion is  that we  encounter  two distinct   dynamical patterns of behavior that  follow equally real (realistic) clocks.
The Euclidean mapping (Wick rotation) is merely a mathematical artifice connecting the pertinent  dynamical   models, or rather transforming one
model into another.

\section{\textbf{Comments on variational extremum principles}}

\subsection{ (Shannon) Entropy extremum principle}

  Given $V=V(x)$, fix a priori   $ \langle V \rangle = \zeta $.
 Extremize
  ${\cal{S}}= -\langle \ln \rho \rangle $  under this  constraint, i.e. seek an extremum of  $$  {\cal{S}}(\rho )  + \alpha  \langle V\rangle  = \langle -\ln \rho   + \alpha  V \rangle  $$
  where $\alpha $ is a Lagrange multiplier.    As an  outcome  we get the    $\alpha $-family of pdfs
 $\rho _{\alpha }= A_{\alpha }  \exp[ \alpha  V(x)]$,  provided $(A_{\alpha })^{-1}= \int \exp[\alpha   V(x)]\, dx$ exists. The Lagrange multiplier
 $\alpha $-value  must be inferred from the constraint   $ \langle V \rangle  _{\alpha }  = \zeta $.

\subsection{Fisher information extremum principle}

Fix a priori   $ \langle V \rangle = \zeta $.  Extremize the Fisher information measure
   ${\cal{F}}(\rho ) $ under that \textbf{constraint}:
  $$ {\cal{F}}(\rho ) +\lambda  \langle V\rangle   = \langle  ({\nabla \ln \rho })^2 +\lambda V  \rangle  $$
  Remember  that   $-  \langle  {\frac{\Delta \rho ^{1/2}}{\rho ^{1/2}}} \rangle  = {\frac{1}4}{\cal{F}}(\rho )$.
    The extremizing pdf $\rho (x)\doteq \rho _*(x)$  comes out from :
  $$
  V(x)=  {\frac{2}{\lambda }}\,   [{\frac{\Delta \rho }\rho }  - {\frac{1}2} {\frac{(\nabla \rho )^2}{\rho ^2}}] = + {\frac{4}{\lambda }}  \, {\frac{\Delta \rho ^{1/2}}{\rho ^{1/2}}} $$
  Outcome:  $\lambda  $-family of pdfs;   $\lambda $ gets fixed by  $\langle V\rangle _{\lambda }= \zeta $.
  Setting $\lambda  = 2/mD^2$, we recover the Brownian framework;
   $\lambda = 8m/ \hbar ^2$  is admitted as a special case.

\subsection{Hamilton-Jacobi route}

Think of a purey classical case  $H=p^2/2m +V(x)$, \{$\dot{q}=p/m$, $\dot{p}= -\nabla V (q)$\}. Next, assign  random initial data (here, in space)
 $\rho _0(x)$  $\Rightarrow {\cal{S}}(\rho )$
and ${\cal{F}}(\rho )$. By extrmizing the action functional we deduce the standard Hamilton-Jacobi description of an ensemble of classical systems:
\begin{equation}
I_0(\rho , s)= \int_{t_1}^{t_2}  \langle \left[ \, \partial _ts  +{\frac{1}{2m}} (\nabla s)^2   +V\right] \rangle \, dt  \Longrightarrow
  \partial _t s + {\frac{1}{2m}} (\nabla s)^2   +V =0
  \end{equation}
plus the continuity equation  $\partial _t\rho = - \nabla  (v \rho )$. Here  an assumption
 $v= (1/m) \nabla s$   implies  $\partial _t v + (v\nabla v) =-\nabla V$.

\subsection{Constrained  Fisher information}
   Fix a priori
$\int_{t_1}^{t_2} {\cal{F}}(\rho )(t) \, dt =\zeta $.  Extremize
$$I_{\gamma }(\rho ,s) =   \int_{t_1}^{t_2} dt\, \langle  \left[ \partial _t s + {\frac{(\nabla s)^2}m}  \pm  V \right]  + \gamma  {\frac{(\nabla \rho )^2}{\rho ^2}}   \rangle   \Longrightarrow
 $$
\begin{equation} \partial _t \rho = - \nabla (v \rho )
\end{equation}
$$\partial _t s + {\frac{(\nabla s)^2}m}  \pm V   + 4\gamma \, {\frac{\Delta \rho ^{1/2}}{\rho ^{1/2}}} =0  $$
where  by denoting   $\pm V$ we intend to make a distinction between confining (generically bounded from below)
and scattering potentials.\\
Outcomes  (an admissible case of $\gamma =0$ is left aside):\\
(i) $\gamma   = - mD^2/2$, eventually  followed by setting $D=\hbar /2m$,  leads  to the   $D$-labelled
quantum hydrodynamics  (before, we have referred to $+V$ only)
$$\partial _ts  +   {\frac{1}{2m}} (\nabla s)^2 \pm V +  Q   = 0 $$
(ii)  $\gamma   = +  mD^2/2$, with  the potential term  $-V$ only, leads to the  Brownian hydrodynamics
$$\partial _t s + (1/2m) (\nabla s)^2  - (V + Q)=0 $$
Note:  $t\rightarrow it$ relationship can be  secured  for $+V$, where $V$ is a confining  potential.
$$\partial _ts  +   {\frac{1}{2m}} (\nabla s)^2 + (V +  Q)   = 0 $$
 c.f.   $t \rightarrow it$ $\Longrightarrow $  $\exp(-t\hat{H}/2mD)\Psi _0 = \Psi _t   \longrightarrow \exp(-it\hat{H}/2mD)\psi _0 = \psi _t $  issue.\\
 We demand  $\hat{H}$ to have  a   bottom eigenvalue equal zero  (to yield a contractive semigroup).
 For a bounded from below Hamiltonian this   can be always  achieved,  like e.g. in  case of $\hat{H}= (1/2)(-\Delta + x^2  - 1)$.\\

\subsection{Hamilton-Jacobi route - a  catalogue of "standards"}
In below we emphasise a relevance of the sign of the external  potential (positive - confinement,
 negative-scattering) in Lagranagian desnities.\\
 \noindent
(i)    $ {\cal{L}}^{+}  =  - \rho \left[ \, \partial _ts
      + (m/2)(v^2 + u^2)  +  V\right] $ $\Longrightarrow $   $\partial _t s + (1/2m) (\nabla s)^2  + (V + Q)=0$
\vskip0.2cm
\noindent
(ii)   $   {\cal{L}}^{\pm }_{cl}  =  - \rho \left[ \, \partial _ts
      + (m/2)v^2     \pm  V \right] $  $\Longrightarrow $ $\partial _t s + (1/2m) (\nabla s)^2  \pm V =0$
\vskip0.2cm
\noindent
(iii)  $ {\cal{L}}^{-} =  - \rho \left[ \, \partial _ts
      + (m/2)(v^2  -  u^2) - V\right]  $ $\Longrightarrow $ $\partial _t s + (1/2m) (\nabla s)^2
       - (V + Q)=0$.\\
 A continuity equation $\partial_t \rho = - \nabla (v\, \rho )$ is shared by all listed cases, provided we set
 $v= (1/m) \nabla s$.\\
      On dynamically admitted fields  $\rho (t)$ and $s(x,t)$,   $L(t) =\int dx\, {\cal{L}} \equiv  0$, i.e.
  we have  $\langle \partial _t s\rangle  =- H $.\\
The respective  Hamiltonian functionals have the form:\\
\noindent
(i)  $ H^+ \doteq \int  dx \, \rho \left[(m/2) v^2       +  V   + (m/2) u^2  \right]>0$, is a
 (quantum) constant of motion
\vskip0.2cm
\noindent
(ii) $H_{cl}^{\pm} \doteq  \int  dx \, \rho \left[(m/2) v^2    \pm  V  \right] = E$, \,   $E = (p^2/2m) \pm V(x)$,  constant on each path
\vskip0.2cm
\noindent
(iii) $H^-  \doteq \int  dx \, \rho \left[(m/2) v^2   -  V   -(m/2) u^2  \right] =0 $, identically
 in Brownian motion.\\
We  emphasize   that, from  the start,
  $V(x)$  is chosen  to be  confining. A class of continuous and bounded from below
   functions allows to secure  $\hat{H}\geq 0$. Eventually,  after subtracting the lowest eigenvalue of  the
   bounded from below  energy operator.

\section{Kinetic theory lore: Brownian  analogies and hints.}

 Consider free  phase-space Brownian motion  in the large friction  regime.
   $W(x,u,t)$ stands for  phase-space (velocity-position) probability distribution  with suitable  initial  data at $t=0$. Denote  $w(u,t)$ and $w(x,t)$, the
    marginal pdfs. \\
    We set $D = k_BT/m\beta $  and observe that  actually, in the large friction regime,  $w(x,t)$ stays in the vicinity (and ultimately converges to)  of
    the heat kernel  solution $ w(x,t)\sim (4\pi Dt)^{-1/2} \exp (-x^2/4Dt)$  of  $\partial _tw = D \Delta w$. \\
     We introduce moments and local moments of
     the pdf $w(x,t)$  in the large friction regime:    $\langle u\rangle = \int du\, u \, W(x,u,t)  \rightarrow \langle u\rangle  = (x/2t) w(x,t)$,
  $\langle u\rangle _x =   \langle u\rangle / w(x,t)= x/2t = - D (\nabla w)/w$,
 $\langle u^2 \rangle _x =   \langle u^2 \rangle / w(x,t)= (D\beta - D/2t) +  \langle u\rangle ^2_x $.\\
  The Kramers-Fokker-Planck equation
  \begin{equation}
  \partial _tW + u\nabla _x W = \beta \nabla _u(Wu) + q\Delta _u W
  \end{equation}
   with $q=D\beta ^2$, implies the local conservation laws
  \begin{equation}
    \partial _t w + \nabla (\langle u\rangle  _x  w) =0
   \end{equation}
   $$ \partial _t (\langle u\rangle _x w) + \nabla _x (\langle u^2\rangle _x w) = - \beta \langle u\rangle _x w$$
  Introducing  the  kinetic pressure  notion  $P_{kin}(x,t)= [\langle u^2\rangle _x - \langle u\rangle _x^2] w(x,t)$  we arrive at
   \begin{equation}
    \partial _t +  \langle u\rangle _x\nabla )\langle u\rangle _x  =  - \beta \langle u\rangle _x
    - \nabla P_{kin}/w  \, .
    \end{equation}
  In the large friction regime we have
 \begin{equation}
  -{\frac{\nabla P_{kin}}{w}} =  + \beta  \langle u\rangle _x  - {\frac{\nabla P_{osm}}{w}}
  \end{equation}
  where $P_{osm} = D^2 w \Delta \ln    w$ we name an osmotic pressure of the Brownian motion.\\
We note that  $\nabla P_{osm} = -  w\, \nabla Q/ m$  with
 $Q = -2mD^2 {\frac{\Delta w^{1/2}}{w^{1/2}}}$. \\
Actually   $-  \nabla P_{osm} = (D/2t) \nabla w$.     Thus, denoting $\langle u\rangle _x = v(x,t)$ we arrive at:
  \begin{equation}
  (\partial _t + v\nabla )v = - {\frac{\nabla P_{osm}}{w}} = + {\frac{1}m} \nabla Q
  \end{equation}
  to be compared   with the  general Brownian hydrodynamics  result
   \begin{equation}
   \partial _t v + (v \nabla v) = + {\frac{1}m} \nabla (V+Q)
   \end{equation}
In the past (1992) I have  named all that: "derivation of the quantum potential   from realistic Brownian  particle motions".

\subsection{Functioning of  pressure terms  $P_{kin}$ and $P_{osm}$}

 In view of  $-    \langle \Delta \ln \rho \rangle = {\cal{F}}(\rho )>0$,  the  osmotic  pressure
  $P_{osm}$  is predominantly negative-definite.  To the contrary, the kinetic pressure
   $P_{kin}$ is positive definite. That imposes limitations on the validity of
    the large friction regime, to become  operational after  times  $t> (2\beta )^{-1}$. \\
      Let us introduce the  notion of kinetic temperature:
      \begin{equation}
0 \leq \Theta _{kin} = m {\frac{P_{kin}}{w}}\sim  (k_BT - {\frac{mD}{2t}})< k_BT
\end{equation}
whose (large time limit) asymptotic  value,  $k_BT$ actually  is.
 Since $P_{osm}/w = D^2 \Delta \ln w =  -D/2t$, we learn that  a (predominantly) positive-definite quantity
\begin{equation}
\Theta _{osm} =  - m {\frac{P_{osm}}{w}}  = - mD^2 \Delta \ln w  \Longrightarrow  \Theta _{kin}
\sim  (k_BT    - \Theta _{osm})
\end{equation}
 gives account of the deviation from  thermal  equilibrium,  in terms of  the local  "thermal energy" (agitation)
   $\Theta _{osm}$.\\
One more  useful identity (not an independent equation) is  here  valid. It   expresses the "thermal energy"
 conservation  law
(observe that no  thermal currents are hereby induced):
\begin{equation}
(\partial _t   + v\nabla )  \Theta _{osm}  =  - 2(\nabla v) \Theta _{osm}  \Longrightarrow
 \partial _t \Theta _{osm}  =  - 2(\nabla v) \Theta _{osm}
 \end{equation}

\subsection{Meaning of the pressure term in Brownian  hydrodynamics ($P_{osm}\doteq P$)}
We come  back to  local conservation laws of the Brownian "fluid". This is an ensemble picture
 of the Brownian  motion:  imagine
the  Pablo Picasso art of placing one upon another hundreds of transparent  foils, each carrying  a  drawing of
 one complete  Brownian trajectory. All random  paths are suposed to  start from the same point and  next
allowed to   run   a  pre-defined  time  period  $[0,T]$, common for all repetitions.
Brownian hydrodynamics is about statistical properties of  such  an ensemble:
\begin{equation}
\partial _t v + (v \nabla v) = + {\frac{1}m} \nabla (V+Q)={\frac{1}m}F- {\frac{\nabla P}{w}}
\end{equation}
$$
- {\frac{\nabla P}{w}} = + {\frac{1}m} \nabla Q; \, \, \,   F\doteq   - \nabla (-V)  $$
In normal liquids the pressure is exerted upon any control volume (here-by an imagined  small droplet),
  thus involving  its compression.  Just to the contrary,  in
case of Brownian motion  we deal with a definite  decompression. \\
Let us consider  a reference volume
(control interval, finite droplet) $[\alpha ,\beta ]$ in $R^1$ (or
$\Lambda \subset R^1$) which at time $t>0$ comprises a
certain fraction of particles (it is a  loose term designating, whatever they would be, the
Brownian "fluid" constituents). \\
 The  time rate  of particles loss  or gain  by the volume $[\alpha,\beta
]$ at time $t$, is equal to the flow outgoing through the
boundaries
$$ -\partial _t \int_{\alpha }^{\beta }\rho
(x,t)dx = \rho (\beta ,t)v(\beta ,t) - \rho (\alpha ,t)v(\alpha
,t)$$
To analyze the momentum balance, let us  slightly
deform  the boundaries  $[\alpha ,\beta ]$ to   compensate  the mass imbalance: $[\alpha
,\beta ] \rightarrow [\alpha +v(\alpha ,t)\triangle t,\beta
+v(\beta ,t)\triangle t] $.  Effectively, we pass  to a  locally co-moving
 (droplet) frame; that is the  Lagrangian picture.\\
(i)  The mass balance has been thus established  in  the   moving  droplet:
$$ lim_{\triangle t\downarrow 0} {1\over {\triangle
t}}\bigl [\int _{\alpha +v_{\alpha }\triangle t}^{\beta
+v_{\beta}\triangle t} \rho (x,t+\triangle t)dx - \int_{\alpha
}^{\beta } \rho (x,t)dx\bigr ] =0$$
(ii)  For local matter  flows
$(\rho v)(x,t)$,   in view of  $\partial _t(\rho
v)=-\nabla (\rho v^2) +  (1/m)\rho \nabla (V+ Q)  $, the time rate of
change of momentum  (per unit of mass) of the  droplet, reads
$$ lim_{\triangle
t\downarrow 0}{1\over {\triangle t}}\bigl [\int_{\alpha +
v_{\alpha }\triangle t}^{\beta +v_{\beta }\triangle t} (\rho
v)(x,t+\triangle t)- \int_{\alpha }^{\beta } (\rho v)(x,t)\bigr ]=
\int_{\alpha }^{\beta } \rho  {\frac{1}m}  \nabla (V+Q) dx  $$
\centerline{However, $\nabla Q /m = -  {{\nabla P}\over \rho }$ and $P=D^2\rho
\triangle ln \rho $. Therefore:}
$$ \int_{\alpha }^{\beta }\rho \,
{\frac{1}m} \nabla (V+Q)dx = \int_{\alpha }^{\beta } \rho \nabla \Omega
dx - \int_{\alpha }^{\beta } \nabla P dx = {\frac{1}m} E[\nabla V ]_{\alpha }^{\beta } + P(\alpha ,t) - P(\beta ,t) $$
(iii)  The time rate of  change of the  kinetic energy  of   the  droplet is:
$$ lim_{\triangle t\downarrow
0}{1\over {\triangle t}}\bigl [\int_{\alpha + v_{\alpha }\triangle
t}^{\beta +v_{\beta }\triangle t} {1\over 2} (\rho
v^2)(x,t+\triangle t)- \int_{\alpha }^{\beta } {1\over 2}(\rho
v^2) (x,t)\bigr ]= \int_{\alpha }^{\beta } {\frac{1}m} (\rho v) \nabla (V +Q) dx $$
Note that  $\int_{\alpha }^{\beta }  \rho v\nabla Qdx = -
 \int_{\alpha }^{\beta }v\nabla Pdx$ (c.f. the  standard  notion of power
 release ${{dE}\over {dt}} = F\cdot v$)

 \subsection{Meaning of the pressure term in quantum   hydrodynamics ($- P_{osm}\doteq P$)}
We do not care about a specific "trajectory" (mis  or to the contrary)-representation of the quantum motion and pass directly to
local  conservation laws:
\begin{equation}
\partial _t v + (v \nabla v) = - {\frac{1}m} \nabla (V+Q)= {\frac{F}m} - {\frac{\nabla P}{\rho }}
 \Longrightarrow
 \end{equation}
$$     - {\frac{1}m} \nabla Q =    + {\frac{\nabla P_{osm} }{\rho }} \doteq  -  {\frac{\nabla P }{\rho }} $$
which enforce  $- P_{osm}= - D^2 \rho  \Delta \ln  \rho   \doteq P$, $D=\hbar /2m$, while $F=-\nabla V$.
 If compared to the
 Brownian hydrodynamics all  $(V+Q)$ contributions come with an inverted sign. This  carries over to the
  mass, momentum and  kinetic  energy rates.\\
Quite at variance with  the Brownian $P=  P_{osm}$, the quantum pressure term $P= -P_{osm}$
is predominantly positive.
 We recall that $-\langle \Delta \ln \rho \rangle = \langle
{\frac{(\nabla \rho )^2}{\rho ^2}}\rangle = {\cal{F}} (\rho )>0$.  \\
We note in passing that   quantum mechanically derivable heat  transfer equation
\begin{equation}
(\partial _t + v\nabla )\Theta _{osm}  = -2{\frac{\nabla q}{\rho }}   - 2(\nabla v) \Theta _{osm}
\end{equation}
with $\Theta _{osm} =  - m {\frac{P_{osm}}{\rho }}  = - mD^2 \Delta \ln \rho  $ and $q= -2mD^2 \rho \Delta v$,
 reproduces the   Brownian form,  at least  for generic free Schr\"{o}dinger wave packets
with  $\Delta v =0$. We get $\partial _t \Theta _{osm}  =  - 2(\nabla v) \Theta _{osm}$  as well.
There is no heat current in such case.

\section{Hamilton-Jacobi related hydrodynamics  and  (Bohmian) trajectory descriptions.}

At this poit it seems instructive to make a comment on uses of  the Eulerian picture (passive control)  vs
Lagrangian picture (active control in a co-moving frame)  of hydrodynamical equations of motion.
Let us  simply  give our droplet, previously considered  as a co-moving control volume,
 an infinitesimal size.  We readily  identify  the (fairly small sized) droplet dynamics: each droplet
  "looks"  particle-like,  while  following
  Bohm-type trajectories.\\
  Since,  $f(x,t) \rightarrow f(x(t+ \Delta t), t+ \Delta t) \sim  [\partial _t  f + (v\nabla) f ]\Delta t;   \, \, \, \, \, \,
\dot{x}= v= v(x,t)_{|x(t)=x}$, with
$x(t+\Delta t) \sim v \Delta t$, $v= (1/m) \nabla s$   and
$\partial _t s  =   {\frac{ds}{dt}} - m v^2$,  we  are  in fact bound to  work with:\\
(i)   Classical  hydrodynamics: (droplet) paths in the Lagrangian frame
\begin{equation}
 {\frac{d\rho}{dt}} = - \rho  \nabla v   \longrightarrow  \rho (x(t+\Delta t),t +dt)
  \sim \exp[ - (\nabla v)\Delta t] \, \rho (x,t)
  \end{equation}
  $$ {\frac{ds}{dt}} = {\frac{1}{2m}}  (\nabla s)^2 -  (\pm V) \Longrightarrow
m{\frac{dv}{dt}} = - \nabla (\pm V)$$
(ii)   Brownian  hydrodynamics:  (droplet)  paths   in the Lagrangian frame
\begin{equation}
 {\frac{d\rho}{dt}} = - \rho  \nabla v
 \end{equation}
 $$ {\frac{ds}{dt}} = {\frac{1}{2m}}  (\nabla s)^2 + (V + Q) \Longrightarrow
m{\frac{dv}{dt}} = +  \nabla (V +Q)$$
We  need to  recall a purely random (Wiener noise)  background of the hydrodynamical formalism.  We
encounter  here  a primordial description in terms of  random  variables and  paths, The latter
may cross the droplet (e.g.  enter from the outside, leave or
simply stay within  for a while):
 $dX(t) = b(X(t))dt + \sqrt{2D} dW(t) \Longrightarrow
\partial _t\rho = D\Delta \rho - \nabla (b\rho ); \, \, \, \,  {\frac{V(x)}{2mD}} =
  mD\left[ {\frac{b^2}{2D}} + \nabla b\right] $ \\
(iii)   Quantum   hydrodynamics:  (droplet)  paths   in the Lagrangian frame $\Longrightarrow $ Bohmian  trajectories
\begin{equation}
{\frac{d\rho}{dt}} = - \rho  \nabla v
\end{equation}
$$ {\frac{ds}{dt}} = {\frac{1}{2m}}  (\nabla s)^2 - (V + Q) \Longrightarrow
m{\frac{dv}{dt}} = - \nabla (V +Q)$$

\section{Acceleration concept in  random motion: $II^{nd}$ Newton law}
\vskip0.2cm
Consider a Markovian diffusion process on $R$, for times $t \in [0,T]$:  $dX(t) = b(X(t),t)dt +  \sqrt{2D} dW(t)$,
where $W(t)$ stands for the Wiener noise  and $X(t_0) =x_0$.   Given  $p(y,s,x,t), s\leq t$  and
$\rho _0(x)$, we  can infer  a  statistical future of the process:
\begin{equation}
  \rho (x,t) = \int \rho (y,s) p(y,s,x,t) dy
\Longrightarrow \partial _t\rho = D\Delta - (\nabla b\rho )
\end{equation}
$$
b(x,t) = \lim_{\Delta \rightarrow 0} {\frac{1}{\Delta t}} \int (y-x) p(x,t,y,t+\Delta t)dy =  v(x,t) + (D\nabla \rho /\rho )(x,t)$$
We can  as well  reproduce a statistical past  of the process, by means of

\begin{equation}
p_*(y,s,x,t) \doteq p(y,s,x,t) {\frac{\rho (y,s)}{
\rho (x,t)}}\Longrightarrow \rho (y,s) = \int p_*(y,s,x,t)\rho (x,t)dx
\end{equation}
$$b_*(y,s) =\lim _{\Delta s \rightarrow 0} {\frac{1}{\Delta s}} \int (y-s)  p_*(x,s - \Delta s,y,s)dx = v(y,s) - (D\nabla \rho /\rho )(y,s)$$
Making notice of  $v= (1/2)(b+b_*)$, we get:
\begin{equation}
\partial _t\rho = -\nabla (v\rho )= D\Delta \rho - (\nabla b\rho ) =
 - D\Delta \rho - \nabla (b_* \rho ).
 \end{equation}
Consider  $b= DX$ and  $b_* = D_*X$  as special cases of forward (predictive) and backward (retrodictive)
 time derivatives  of functions of the random variable $X(t)$:\\
\begin{equation}
(Df)(X(t),t) = (\partial _t +b\nabla + D\Delta ) f(X(t),t)
\end{equation}
$$(D_*f)(X(t),t) = (\partial _t  + b_*\nabla -  D\Delta )f(X(t),t)$$
Analyze  acceleration  formulas  for  diffusion-type processes, in terms of the
 time rate of change (forward and backwards) drift functions and their local averages.
 Their response to small time increments reads:\\

\noindent
 (i) $b(x,t) \rightarrow b(X(t +\Delta t),t + \Delta t)$ $ \Longrightarrow $
 $\langle  b\rangle (x,t+\Delta t)$\\
\, \, \, \, $ \langle b \rangle (x,t+\Delta t) = \int p(x,t,z,t+\Delta t) b(z,t+\Delta t) dz \sim  b(x,t) +
 (D^2X)(t)\Delta t$.\\
(ii) $b(X(t-\Delta t),t-\Delta t) \Longrightarrow
 \langle b \rangle (x,t-\Delta t) \rightarrow b(x,t)$\\
\, \, \, \, $\langle b \rangle (x,t-\Delta t) =  \int b(y, t-\Delta t)p_*(y,t-\Delta t,x,t) dy \sim
b(x,t) - (D_*DX)(t) \Delta t$.
\\
(iii) $b_*(x,t) \rightarrow   b_*(X(t+\Delta t),t+\Delta t) \Longrightarrow
\langle b_*\rangle (x,t+\Delta t)$.\\
\, \, \, \, $\langle b_*\rangle (x,t+\Delta t)  \sim  b_*(x,t) + (DD_*X)(t)\Delta t $\\(iv)  $b_*(X(t-\Delta t),t-
\Delta t) \Longrightarrow \langle b_*\rangle (x,t-\Delta t)   \rightarrow  b_*(x,t)  $
$\langle  b_*\rangle (x,t-\Delta t)   \sim   b_*(x,t) - (D_*^2X)(t)\Delta t$. \\

Accordingly, we can associate various acceleration
 formulas with general   diffusion-type processes.  They may be interpreted as stochastic analogues of the
${II}^{nd}$ Newton law in the (local) mean: accelerations are related to conservative volume forces of external
origin.\\
We  indicate that  by over-emphasizing the   Newtonian  viewpoint,  we get  somewhat blured
  an important  intrinsic   acceleration input,  due to  the  background random motion.
    It is encoded  in the   $\pm {\frac{1}m} \nabla Q$ contribution  to the current velocity
     time rate of change.  \\
Following the standard  association of accelerations with external (volume) forces, one is inclined
to attribute the second Newton law meaning to the formula
\begin{equation}
(D^2X)(t) =  (\partial _t + v\nabla )v  - {\frac{1}m} \nabla Q  = (D^2_*X)(t) =
  + {\frac{1}m} \nabla V
\end{equation}
  in case of  the Brownian motion, while
\begin{equation}
{\frac{1}2} [(DD_* + D_*D)X](t)= (\partial _t + v\nabla )v  + {\frac{1}m} \nabla Q = - {\frac{1}m}\nabla V
\end{equation}
in case of  Nelson's stochastic mechanics, i.e.  the  diffusion-type  probabilistic
 counterpart of   the  Schr\"{o}dinger picture quantum motion.

It is the mid-term in the  formulas (49) and (50),  where the   role   of $\mp \nabla Q$    must be strongly emphasized. Set   $\nabla V=0$;
 the mid-term   still  accounts for definite acceleration phenomena,  that are intrinsic to the random motion proper  (c.f. a discussion of Section 5):\\
\begin{equation}
 (i)\, \,  {\frac{dv}{dt}} -
{\frac{1}m}\nabla Q =0, \, \, \, \, (ii)\, \,   {\frac{dv}{dt}} +  {\frac{1}m}\nabla Q=0
\end{equation}
Therefore, as appropriate candidates  for the   ${II}^{nd}$ Newton law, we promote  not (47) or (48), but rather  the  (hydrodynamical)
local conservation laws:
\begin{equation}
(iii)\, \, (\partial _t + v\nabla )v =   \pm {\frac{1}m} \nabla  (Q +  V)
\end{equation}
where the (seemingly minor) sign difference of the right-hand-side terms is hereby  crucial and will set grounds
in below to  the $III^{rd}$ Newton law in the local mean.

\section{$III^{rd}$ Newton law}
\subsection{Impulse-momentum change law for for small times}
Both the acceleration and impulse-momentum change concepts  have been borrowed  directly from  classical
mechanics. Nonetheless,  an exploitation of  properly tailored   local mean values allows  to  extend the
  meaning  of   purely  mechanical  concepts  to
  the theory of random motion (e.g. diffusion-type  processes, with  continuous, but  generically
non-differentiable  sample paths). \\
That was the case in connection with  the $II^{nd}$ Newton law. It could have been  explicitly rooted
in the  impulsive  (short time increments)  behavior of drifts in the  Brownian motion:
 \begin{equation}
  b_*(x,t) - \langle  b_*\rangle (x,t-\Delta t)  \sim  \langle  b\rangle (x,t+\Delta t) - b(x,t)
   \sim  {\frac{1}m} \nabla V \,   \Delta t
   \end{equation}
and/or an  impulsive behavior of drifts  in  stochastic mechanics (probabilistic counterpart of
the Schr\"{o}dinger picture evolution)
\begin{equation}
 b_*(x,t) - \langle  b_*\rangle (x,t-\Delta t)  \sim  \langle  b\rangle (x,t+\Delta t) - b(x,t)
   \sim  {\frac{1}m} \nabla (V  + 2Q) \,   \Delta t \, .
 \end{equation}
However, there is  no other than esthetic  reason to associate the  Brownian acceleration with the first and not
the second (stochastic mechanics) formulas above. Indeed, more careful examination  proves  that
 the Brownian motion can be characterized   on an equal  footing by  both acceleration definitions:
\begin{equation}
 (D^2X)(t)= (D^2_*X)(t) =   + {\frac{1}m} \nabla V  \Longleftrightarrow
{\frac{1}2} [(DD_* + D_*D)X](t)={\frac{1}m}\nabla (V +2Q)
\end{equation}
and  likewise, in the  stochastic  description of quantum motion (e.g.  stochastic mechanics):
\begin{equation}
 (D^2X)(t)= (D^2_*X)(t) =  - {\frac{1}m}\nabla (V +2Q)  \Longleftrightarrow
{\frac{1}2} [(DD_* + D_*D)X](t)= - {\frac{1}m}\nabla V
\end{equation}
It is now clear that the Brownian motion and  the stochastic transcription of quantum motion  (stochastic
mechanics) do differ  fundamentally  in their response to both random noise and external forces.
 We may stay happy with a  that observation, e.g. a  clearly identified difference between two types of
  random motion.  Accordingly, they   can  be interpreted  as totally divorced  from each other,
   except for incidental formal connections (c.f. the Euclidean time and the  Wick transformation  issue).

\subsection{Action-reaction rule}

There is still another possibility, that may be  given a physical status of relevance,
albeit  on  sufficiently small time-scales only.
Considering the  impulse-momentum change law as a valid property of random motion,
 we may  as  well   interpret the  two  considered   dynamical patterns of behavior as
being    involved in  a  perpetual   Brownian -- anti-Brownian  acceleration  intertwine.
That,  via the  ${III}^{rd}$ Newton law in the mean. In the past  (1992, 1999)we have  heuristically  formalized this
 idea in the concept of the {\it  Brownian recoil principle.\rm }

 Indeed,   all acceleration expressions in   Eqs. (55)    can be mapped into  those of Eqs. (56), and in reverse,
  by addition/subtraction  of the force term:
 \begin{equation}
\Downarrow   \pm {\frac{2}m} \nabla (V+Q) \, \, \, \,
  \Uparrow
  \end{equation}
  provided, at a  given time instant we  per force  attribute the same values of $\rho(x,t)$ and $v(x,t)$
   to both the Brownian and quantum  hydrodynamical fields.

    If the two motions are coupled by the action-reaction
   principle ($III^{rd}$ Newton law) at time $t$, then  an impulse - momentum (here, velocity field)
   change  law can be used to predict  $+\Delta t$ updated values of $\rho $ and $v$.
  Thus e.g. the  Brownian impulse in a co-moving frame (given $\rho $ and $v$) induces
\begin{equation}
\Delta \rho =  -[(\nabla v)\,\Delta t]\, \rho   \, \, \, \, \, \, \, \,
m\Delta v = + \nabla (V+Q)\, \Delta t
\end{equation}
while an accompanying anti-Brownian  impulse in a co-moving frame (given $\rho $ and $v$)
reads
\begin{equation}
\Delta \rho =  - [(\nabla v)\,\Delta t]\, \rho  \, \, \, \, \, \, \, \,
m\Delta v = - \nabla (V+Q)\, \Delta t \, .
\end{equation}
We do not attempt to address the celebrated "egg-before-hen" dilemma.   The formulas (58) and (59) are regarded
as a statistically relevant   record of   the {\it  recoil  \rm}  effect  where
an "anti-Brownian" impulse associated with   a quantum particle (we do not bother what is  actually  meant
 under this  notion) induces,    and in turn gets induced,    by  the  Brownian   motion  pulse  excited   in  a  dissipative  random medium
   (any conceivable  notion of a  surrounding medium,  like e.g.  the  "vacuum", zero point radiation field etc.).
The dissipation  assumption  destroys the symmetry of the action-reaction picture, since the Brownian pulse
should quickly decay.

 \subsection{A concept of the Brownian recoil principle}

 Consider  $\Delta t \ll 1$. Within $[t,t+\Delta t]$, let  the  action-reaction coupling
 between   the  "vacuum"  (whatever that may be)  and matter particles  sets rules   of
 the game   $\Longrightarrow $
  $\langle \Delta p\rangle _{vacuum} + \langle \Delta p \rangle _{matter} =0$.\\
  \noindent
   The "vacuum turbulence"   propels matter  particles   by transferring them   an    anti-Brownian
 recoil  impulse (set $D=\hbar /2m$),  whose "vacuum" trace  (and reason)  is the   Brownian impulse (may quickly
  die out due to dissipation, we track the matter data).\\

{Step $I$}.  Given the matter data  $\rho(x,t)$ and $v(x,t)$. At $t+\Delta t$ we have
$\rho  + \Delta \rho =  \exp [-(\nabla v)\, \Delta t]\, \rho $ and
  $v\rightarrow v+ \Delta v$, where  the action  ("vacuum" impulse)
  \begin{equation}
\Delta v = +  {\frac{1}m} \nabla (V + Q) \, \Delta t \,  \,  \, \, (Brownian)
\end{equation}
is paralleled by   the reaction  (matter impulse): ($\Downarrow $ - subtract; $\Uparrow  $ - add:  ${\frac{2}m}\nabla  (V+Q)$ \,  !)
\begin{equation}
\Delta v   = - {\frac{1}m} \nabla (V +  Q)\, \Delta t \, \, \, \,  (anti-Brownian, e.g.\, \,   quantum)
\end{equation}

{ Step $II$}.  Update the matter data to $\rho(x,t+\Delta t)$, $v(x,t+\Delta t)$,  leave aside
those referring to the "vacuum"
and to the preceding Brownian impulse, turn to the next $\Delta t$ episode when both impulses are
excited anew. The  new (updated) values of $\rho $ and $v$ at time  $t+\Delta t$ are presumed to
be determined by the  anti-Brownian impulse again.

 Any  physical  justification of the  Brownian recoil principle  needs a double-medium picture:\\
 (i)  an  active "vacuum" (background random field, non-equilibrium reservoir, zero-point fluctuations)
   that is  generating and  supporting    Brownian  pulses.  These may  be interpreted in terms
   virtual particles \\
 (ii) matter particles, whose dynamics is governed by the $III^{rd}$  Newton law and the resultant
  recoil effect.\\
A detailed   theory of  the  "vacuum"-particle coupling is obviously necessary  to go beyond heuristics.

With an inspiration coming from Yves Couder's lecture on particle-wave associations,
 and from  deepened studies of the role of  trajectory descriptions in difraction/intererence phenomena,
 as a residual  subfield of   quantum chemistry, we end up with a  simple
  statement that  there is plenty of room down there.
Indeed,   before going into any sophisticated formal arguments,  one should  always keep in mind  the
 quantum mechanical scales of interest:
 atomic nucleus size  $\sim 10^{-15}- 10^{-14}m$,  atom size
 $\sim  10^{-10}-10^{-9}m$,  electron size (whatever that means for inspired theoreticians) $\sim 10^{-15}m$,
 possibly down to $\sim 10^{-18}m$.\\
Presumably it is not devastatingly naive to address na issue  of the   (Schr\"{o}dinger's wave function)
  $\psi $-ness   of the   electron "cloud" in  the atom,  while realizing that, with or without the second
  quantization and with or without  quantum
 electro- or  chromodynamics,   the  "vacuum" (not an empty void) functioning in quantum physics is still
  an open territory.

\section{Bibliographic notes}
There is no way to give justice to all contributors in the field of quantum hydrodynamics or
stochastic mechanics.  Our selection of references will be less  then modest and in addition to Nelson's,
Holland's and Wyatt's contributions, will mainly concentrate  on  a sample my own research in this area.
More references (with  a bibliography of the subject) can be found and retrieved from my
personal Web page:  http://www.fiz.uni.opole.pl/pgar/.

My own hunch is that the Schr\"{o}dinger picture quantum motion admits a consistent representation
 in terms of diffusion-type processes. That was the main idea of Nelson's stochastic  mechanics.  However,
 in search for "reasons of randomness"  we have found unavoidable  to admit a coupled  double-medium picture.

   Its  qualitative features seem to be not distant from the particle-wave association picture
 (with all reservations raised in the original paper due to Y.  Couder and E. Fort,  2006), in which
  a  physically "real"  particle  induces a  physically relevant  "realistic"  wave,
   and  that  wave in turn  is capable of  affecting  further  motion of the particle.   A conceptual
   input of diffusion waves might be useful at this point,   as well.

 \end{document}